%% file: main.tex
\documentclass[twocolumn,twocolappendix,nofootinbib]{openjournal}
\usepackage{mathptmx}
\usepackage[T1]{fontenc}
\DeclareRobustCommand{\VAN}[3]{#2}
\let\VANthebibliography\thebibliography
\def\thebibliography{\DeclareRobustCommand{\VAN}[3]{##3}\VANthebibliography}
\usepackage{graphicx}	% Including figure files
\usepackage{amsmath}	% Advanced maths commands
\usepackage{amssymb}	% Extra maths symbols
\usepackage{rotating}
\usepackage{sidecap}
\usepackage{float}
\usepackage{stfloats}

\usepackage{aecompl}
\usepackage{array}
\usepackage{times}
\usepackage{url}
\usepackage{xcolor}
\usepackage{soul}
\usepackage{orcidlink}

\usepackage{hyperref}
\hypersetup{colorlinks=true,citecolor=blue,linkcolor=blue,filecolor=blue,urlcolor=blue}
\usepackage{orcidlink}

\newcommand{\msun}{{\,\rm M_\odot}}

\begin{document}
\journalinfo{The Open Journal of Astrophysics}

%\date{Accepted ???? Received ????; in original ????}

%\pubyear{2023}

%\maketitle

\title[Projection effects on the splashback feature]{The effects of projection on measuring the splashback feature}

\author{
Xiaoqing Sun$^{1}$ ,
Stephanie O'Neil$^{2,3}$\orcidlink{0000-0002-7968-2088},
Xuejian Shen$^1$\orcidlink{0000-0002-6196-823X}, and
Mark Vogelsberger$^{1,4}$\orcidlink{0000-0001-8593-7692}
}
\thanks{E-mail: xqsun@mit.edu }
\affiliation{
$^{1}$Department of Physics and Kavli Institute for Astrophysics and Space Research, Massachusetts Institute of Technology, Cambridge, MA 02139, USA\\
$^{2}$Department of Physics \& Astronomy, University of Pennsylvania, Philadelphia, PA 19104, USA\\
$^3$Department of Physics, Princeton University, Princeton, NJ 08544, USA\\
$^{4}$The NSF AI Institute for Artificial Intelligence and Fundamental Interactions, Massachusetts Institute of Technology, Cambridge, MA 02139, USA\\
}

\begin{abstract}
The splashback radius $R_{\rm sp}$ is a boundary of a halo that broadly separates infalling and accreted matter. A commonly adopted proxy for $R_{\rm sp}$ is the radius of the resulting steep drop in the density profile, $R_{\rm st}$. Observationally, $R_{\rm st}$ can be measured through fitting the projected galaxy number density profile of the halo, but there has been some discrepancy between the observed and expected $R_{\rm st}$. Therefore, we investigate whether the projection of the density profile onto the plane of the sky could lead to any systematic bias in determining $R_{\rm st}$, by studying the true 3-dimensional and projected halo density profiles from the IllustrisTNG simulation. We investigate a range of projection lengths, and find that $R^p_{\rm st}$ obtained from projected profiles is close to the true $R^*_{\rm st}$, but has a slight decreasing trend with increasing projection length. We also quantify the prominence of the splashback feature and find how the feature shape changes with projection length.
\end{abstract}

\maketitle

% \begin{keywords}
% methods: numerical -- galaxies: haloes -- galaxies: clusters: general -- galaxies: formation -- cosmology: dark matter -- cosmology: large-scale structure of universe.
% \end{keywords}

\input{introduction}

\input{methods}

\input{results}

\input{conclusions}

\section{Acknowledgements}

We thank the anonymous referee for useful suggestions as well as Bhuvnesh Jain, Susmita Adhikari, and Benedikt Diemer for helpful discussions.
X.S. was supported by the MIT Undergraduate Research Opportunities Program (UROP).
S.O. was supported by the National Science Foundation under Grant No. AST-2307787.
Some of the computations were performed on the Engaging cluster supported by the Massachusetts Institute of Technology.

We made use of the following software for the analysis:
\begin{itemize}
    \setlength\itemsep{-.1cm}
	\item {\textsc{Python}}: \citet{vanRossum1995}
	\item {\textsc{Matplotlib}}: \citet{Hunter2007}
	\item {\textsc{SciPy}}: \citet{Virtanen2020}
	\item {\textsc{NumPy}}: \citet{Harris2020}
\end{itemize}

\section{Data Availability}
The data is based on the IllustrisTNG simulations that are publicly available at \url{https://tng-project.org} \citep{Nelson2019}.
Reduced data are available upon request.

\bibliographystyle{mn2e}
\bibliography{bibliography}

\appendix
\input{appendix_fit}

\end{document}

%% file: introduction.tex
\section{Introduction}
\label{sec:intro}

Galaxies grow and evolve in a variety of environments, from isolated galaxies in the field to dense cluster environments full of hot intracluster gas \citep{Smith2005,Lim2021}.
These environments can greatly impact the evolutionary pathways of the galaxies.
For example, cluster galaxies tend to be quenched due to loss of gas, are older and more elliptical, and do not follow the same scaling relations as galaxies in the field \citep{Dressler1980,Cooper2006,Donnari2021,O'Neil2024}. However, haloes are not clearly defined objects but instead refer to generally overdense regions within a cosmic web. It is therefore not straightforward exactly where to draw the boundary separating galaxies within a halo and outside a halo. Understanding the spatial boundaries of dark matter halos is therefore an important question for our understanding of galaxy evolution and large-scale structure.

A halo is commonly defined as a spherical region within which the average mass density is a constant multiple of some reference density, often the critical or mean density of the universe.
The critical density is the total energy density required for a flat universe, $\rho_c = \frac{3 H^2(t)}{8 \pi G}$, where $H(t)$ is the time-dependent Hubble parameter and $G$ is the gravitational constant. The mean density is the matter component of this total density, $\rho_m = \Omega_m \rho_c$. Then, the halo enclosed density is typically defined as 200, 500 or 2500 times the reference density.
The spherical top-hat collapse model \citep{gunn1972} predicts a halo density close to $200$ times the critical density, which is often used to define a halo.

However, since the reference critical density is time-dependent as the Universe expands, the region within the halo boundary must correspondingly change to maintain the required ratio to the reference density. As the Universe expands, $\rho_m$ decreases, so the radius of the halo increases, resulting in a pseudo-evolution of halo size in addition to any physical evolution of the halo \citep{Diemer2013}.

To avoid the issue of pseudo-evolution, it is possible to define a halo boundary based solely on the dynamics of the halo itself.
A more physically motivated definition of a halo would depend on the boundary between infalling and accreted matter. For instance, the boundary that encloses the first orbits of infalling particles, which is where these particles reach their first apocentre, is known as the ``splashback radius'' \citep{Diemer2014, Adhikari2014, More2015}, and is what we focus on in this work. 

The splashback radius can be found in several ways.
The trajectories of dark matter particles can be directly traced in simulations, as in \citet{Diemer_2017}, to determine the first apocentre radii and thus the splashback radius, which encloses these orbits.
As not all particles in a halo will reach the same distance from the halo centre on their apocentre, typically a percentage of these apocentres will be chosen to enclose and define the splashback radius.
However, this is computationally expensive and not feasible observationally.
A ``splashback shell'' can be found instead, as in \citet{Mansfield_2017}.
This method finds the minima of the potential along various lines of sight from the halo centre and creates a boundary along these minima that is not necessarily spherical.

Finally, the most straightforward way to make this measurement is to find the point of steepest slope in the radial density profile of the halo.
\citet{Diemer2014} showed that accreting halos exhibit a steepening in their density profile, and \citet{More_2015} used this steepening in the stacked density profile to find the splashback radius of halos in different mass and accretion rate bins.
It is important to note that there is not a one-to-one matching between the particle trajectories and the point of steepest slope.
Thus, although the point of steepest slope in the density profile is often used as a proxy for the splashback radius, they are not strictly the same quantity. We therefore use the notation $R_{\rm st}$ to refer to the point of steepest slope and $R_{\rm sp}$ to refer to the splashback radius as originally defined enclosing the first orbits of infalling material.

Splashback radii are, by definition, dependent on the 3-dimensional density profiles of halos. However, we observe a 2-dimensional projection of the halo onto the plane of the sky, and it is difficult to reconstruct the full 3-dimensional density profile of halos.
Thus, we rely on the projected density profiles with potential cluster membership contamination to measure the point of steepest slope.
In addition, cluster identification is difficult, as galaxy membership is typically determined probabilistically based on the two sky-plane spatial coordinates and the colour of galaxies \citep[e.g.][]{Koester2007,Rozo2007,Rykoff2014,Klein2019}

There have been several observational studies identifying the steepening slope characteristic of the splashback radius using surveys such as the Sloan Digital Sky Survey (SDSS) and the Dark Energy Survey (DES) \citep[e.g.][]{More_2016,baxter2017,Chang_2018}.
However, observational values tend to underestimate the values predicted from simulations. For instance, in \citet{More_2016}, the measured $R_{\rm st}$ from SDSS is $\sim70-80\%$ of the predicted $R_{\rm st}$ from simulations for both low and high average cluster member distance halos; in \citet{Chang_2018}, the $R_{\rm st}$ obtained from a galaxy density profile in DES is also $\sim 77\%$ that of simulation results.

This can be attributed to many variables, including the biased selection effects in which galaxies are observed and contribute to generating the density profile.
For example, \citet{O'Neil2022} showed that more massive and luminous galaxies trace out a smaller splashback radius than less massive and dimmer galaxies.
Additionally, cluster selection can have a large impact on the measurement.
Studies using SZ-selected clusters \citep[e.g.][]{Shin2019,Zurcher2019,Adhikari2021,Shin2021} find splashback radii that are closer to those predicted compared to studies using optically selected clusters \citep[e.g.][]{More_2016,baxter2017,Nishizawa2018,Murata2020}. Projection along the line of sight could also cause a bias in determining cluster membership, where clusters are contaminated by foreground or background groups \citep{Zu2017,BuschWhite,Sunayama2019}. This in turn alters the shape of the measured density profile. %However, little analysis has been done on how the projection length of the density profile itself could affect the determination of $R_{\rm st}$.

Therefore, we aim to systematically investigate whether the projection of the density profile along the line of sight and inclusion of galaxies within the projected cylinder result in biases in fitting and determining the point of steepest slope.  We use various projection lengths and determine the density profile with the number density of galaxies. With knowledge of the full 3D structure of halos from cosmological simulations, we can compute a ``true'' point of steepest slope. This is compared with the ``observed'' point of steepest slope obtained by projecting the true profiles onto the plane of the sky of an imaginary observer. We focus on using galaxy number density profiles, rather than mass profiles of dark matter or gas particles, as this is a simpler quantity to measure observationally and to isolate the effects of projection. \citet{O_Neil_2021} and \citet{O'Neil2022} showed how the point of steepest slope measured from galaxy number density profiles is generally similar to that measured from dark matter mass density profiles when the full population of galaxies is considered rather than a selected subset.

The rest of the paper is structured as follows. In section 2, we describe how we stack and obtain halo density profiles, then find the true and projected splashback radius. In section 3, we compare projected profiles and splashback radius with their true values, and discuss the effects of projection on the splashback feature. In section 4, we summarize our results.

%% file: methods.tex
\section{Methods}
\label{sec:methods}

\subsection{Simulation}
\label{sec:methods_simulation}
In this work we use IllustrisTNG, a suite of cosmological, magnetohydrodynamical simulations performed using moving-mesh code AREPO \citep{Springel_2010}, with a comprehensive galaxy formation model as described in \citet{nelson2021illustristngsimulationspublicdata}. It uses cosmological parameters consistent with the Planck Collaboration \citep{Planck2016}: $\Omega_m = 0.3089, \Omega_b = 0.0486, \Omega_\Lambda = 0.6911, \sigma_8 = 0.8159, n_s = 0.9667$ and $h=0.6774$.

We use the largest simulation volume TNG300 which is a periodic cube of side length 302 comoving Mpc, and its highest resolution run TNG300-1. We use results from its most recent snapshot at $z=0$.

\subsection{Halo and Galaxy Selection}
\label{sec:methods_selection}
In the simulation, particles are linked with neighbours within a linking length $b=0.2$ times the mean interparticle separation $\sqrt[3]{V/N}$, where $V$ is the volume of the simulation and $N$ is the number of particles, using a Friends-of-Friends algorithm. Gravitationally bound substructures are identified using the SUBFIND algorithm \citep{Springel2001,Dolag2009}. The most massive gravitationally bound object in an FoF group is labelled as a halo, and the others are subhalos.
Each halo has a size $R_{\rm 200m}$ which is the radius of the sphere with enclosed density 200 times the mean density of the universe $\rho_m$. Then the halo mass $M_{\rm 200m}$ is the mass enclosed within $R_{\rm 200m}$. We select halos with $M_{\rm 200m}/\msun \geq 10^{13}$. Galaxies are defined as all subhalos with mass greater than $10^9 \msun$ and non-zero stellar mass. We do not perform additional cuts on galaxies, since this could further bias the splashback radius they trace out \citep{O'Neil2022}, and we want to isolate only the effect of projection on the splashback radius. Since the DM particle and initial stellar particle mass in the simulation are $5.9 \times 10^7 \, \rm M_\odot$ and $1.1 \times 10^7 \, \rm M_\odot$ respectively \citep{firstresultsTNG}, the lowest mass galaxies in our sample would contain roughly 20 particles. Because we are not interested in the resolved properties of the galaxies and only their locations, this is sufficient for our needs. \citet{O_Neil_2021} also showed that a lower resolution does not affect the results of calculating $R_{\rm st}$ using the same galaxy population.

\subsection{Stacking density profiles}
\label{sec:methods_Rsp}
\begin{table}[H]
    \centering
    \begin{tabular}{c|c c c}
        %\hline
        & \rule{0pt}{2.5ex} $\bar{R}_{\rm 200m}$ (kpc) & $\bar{M}_{\rm 200m} (10^{13} \msun)$ & $N_{halos}$ \\ \hline
        \rule{0pt}{2.5ex} $[10^{13},10^{13.5})$ & $792.2\pm87.7$ & 1.70 & 3660 \\ %\hline
        \rule{0pt}{2.5ex} $[10^{13.5},10^{14})$ & $1161\pm130$ & 5.36 & 1291 \\ %\hline
        \rule{0pt}{2.5ex} $[10^{14},10^{14.5})$ & $1684\pm182$ & 16.3 & 390 \\ %\hline
        \rule{0pt}{2.5ex} $[10^{14.5},10^{15})$ & $2444\pm222$ & 49.3 & 65 \\ %\hline
    \end{tabular}
    \caption{In each mass bin, we show the mean mass of halos, radius with standard deviation, and the number of halos.}
    \label{tab:properties}
\end{table}

We aim to find the galaxy number density profile $n(r)$, where $r$ is the radial distance from the halo centre. However, since each individual halo's profile is not spherically symmetric and can be noisy, we stack the profiles of halos with similar masses. 
We choose 4 logarithmically spaced mass bins with edges $M_{\rm 200m}/\msun = 10^{13.0},10^{13.5},10^{14.0},10^{14.5},10^{15.0}$. We summarise the properties of the halos in each mass bin in Table \ref{tab:properties}.

To construct 3D density profiles, we follow these steps:
\begin{enumerate}
    \item For each halo, calculate the distance between all galaxies (ignoring FoF associations) and the halo centre position.
    \item Create 40 linearly-spaced radial bins between $0.01 R_{\rm 200m}$ and $5 R_{\rm 200m}$. Count the number of galaxies that fall into each bin, and divide by the volume of the bin, to create a number density profile for each halo. Normalise the radial coordinate to $R/R_{\rm 200m}$.
    \item Using all the halos within a mass bin, find the mean number density for each (normalised) radial bin to create the stacked profile.
\end{enumerate}

We note that the number of halos stacked is not uniformly distributed between these mass bins as there are many more low-mass halos. However, splitting the low-mass bin only introduced greater error in the stacked profile.

\subsection{Fitting density profiles}
\label{sec:methods_fitting}
For fitting the stacked profiles, we use the density profile in \citet{Diemer2014}. Although the profile is given as a mass density profile $\rho(r)$, we can use the same equation for the number density of galaxies $n(r)$. 
\begin{align}
\rho (r) &= \rho_{\mathrm{inner}} \times f_{\mathrm{trans}} + \rho_{\mathrm{outer}} \label{eq:3D} \\
\rho_{\mathrm{inner}} &= \rho_{\mathrm{Einasto}} = \rho_s \exp{\left( -\frac{2}{\alpha} \left[ \left( \frac{r}{r_s}\right)^\alpha - 1\right] \right)} \\
f_{\mathrm{trans}} &= \left[ 1+ \left( \frac{r}{r_t}\right)^\beta\right]^{\frac{\gamma}{\beta}} \\
\rho_{\mathrm{outer}} &= \rho_{m} \left[ b_e \left( \frac{r}{5 R_{\rm 200m}}\right)^{-S_e} + 1 \right] 
\end{align}
We follow the method in \citet{O_Neil_2021} of fitting the logarithmic derivative of the density profile $\frac{\mathrm{d} \, \log n}{\mathrm{d} \, \log r}$, hereafter called the ``derivative'', which is more robust than other methods such as fitting the profile directly. Specifically, we find the numerical value of $\frac{\mathrm{d} \, \log n}{\mathrm{d} \, \log r}$ using first-order differences, and fit to the analytic expression of $\frac{\mathrm{d} \, \log n}{\mathrm{d} \, \log r}$ using nonlinear least-squares. For a stacked profile, the point where the fitted $\frac{\mathrm{d} \, \log n}{\mathrm{d} \, \log r}$ is minimum corresponds to the normalised splashback radius $R_{\rm st}/R_{\rm 200m}$. The free parameters are $r_s, r_t, \alpha, \beta, \gamma, b_e, S_e, \rho_s$, while $\rho_m$, the background (number) density of galaxies, is fixed at the mean matter density of the Universe divided by  $10^{12}\msun$. Due to degeneracies in the fitting with other parameters, fixing $\rho_m$ should not affect the results significantly. 

The inner profile and parameters $\rho_s$ and $\alpha$ are from the Einasto profile \citep{einasto1965trudy} and capture the scale density and inner slope. The transition term $f_{\rm trans}$ captures steepening ($\gamma, \beta$) around a truncation radius $r_t$. The outer profile is described by the background density $\rho_m$ plus a power law ($S_e$, amplitude $b_e$). We provide some resulting fitted parameters in Appendix \ref{apx:fits}. 

\subsection{Bootstrapping}
\label{sec:methods_bootstrap}
We bootstrap the density profile samples for stacking, to 1. get error estimates on the derivative that improves the fit, and 2. get error estimates for $R_{\rm st}$. For a mass bin with $N$ halos and their corresponding density profiles, we draw an ``outer'' bootstrap sample of size $N$. This is repeated 2048 times. From each ``outer'' bootstrap, we draw another ``inner'' bootstrap of the same size $N$, and compute the stacked profile of those $N$ profiles. We repeat the ``inner'' bootstrap 32 times, to get a mean ``outer'' stacked profile and its error (standard deviation) for each radial bin, which is used to improve the fitting. After fitting, $R_{\rm st}$ is found. Since we have 2048 outer bootstraps, this gives us 2048 values of $R_{\rm st}$, from which we can find a mean and error (16-84th percentile). The values of 2048 and 32 are chosen similarly to \citet{O_Neil_2021}---a smaller number of bootstraps is used for the error bars for fitting, and a larger number of bootstraps for actual estimation of $R_{\rm st}$.

\subsection{Finding $R_{\rm st}$ from projected profiles}
\label{sec:methods_projection}
The methods in the previous sections are used to find the ``true'' 3D profile $n(r)$, and point of steepest slope, which we denote as $R^*_{\rm st}$. To find the projected density profiles, we follow a similar procedure as in the 3D case. 

Taking the $z$-axis as the projection direction (line of sight), the 2D distance between the halo centre and each galaxy is $R = \sqrt{(x-x_{\rm halo})^2 + (y-y_{\rm halo})^2}$, for the ``cylinder'' of galaxies $|z - z_{\rm halo}| \leq d_{\rm max}, R/R_{\rm 200m} \leq 5$ around the halo. This ``cylinder'' of galaxies is simply assumed to belong to that halo. This differs from previous observational studies \citep[e.g.][]{More_2016,Chang_2018} who weight their density profiles by the membership probability of the galaxies. These observational studies also applied magnitude cuts to keep their sample uniform across redshifts, which we do not apply in the simulation.  This makes our sample somewhat more inclusive than what is possible observationally.

For each halo in the simulation, we can take the projection direction to be each of the $x$, $y$ and $z$ axes, to obtain effectively 3 times the number of halos. We stack and bootstrap the density profiles similarly as in the 3D case. The equation for the projected profile is given by integrating the 3D profile along the projection direction. Since we fit using stacked profiles which use the normalized coordinate $R/R_{\rm 200m}$, when integrating we convert $d_{\rm max}$ to the normalized coordinate by dividing by $\bar{R}_{\rm 200m}$ for each mass bin. We note that this integration corresponds to integrating over a cylinder of galaxies around the halo centre, similar to the cylinder we use when computing the observed density profile. 
\begin{equation}
n_{\mathrm{proj,z}}(R) = \int_{-d_{\rm max}}^{d_{\rm max}} n(\sqrt{R^2 + z^2}) \, \mathrm{d}z \label{eq:2D}
\end{equation}

$\frac{\mathrm{d} \, \log n}{\mathrm{d} \, \log R}$ of this equation (the ``derivative'') is fitted to the numerical logarithmic derivative of the stacked projected profiles. The minimum point of this 2D derivative is $R^{\rm 2D}_{\rm st}$. The parameters obtained ($r_s, r_t, \alpha, \beta, \gamma, b_e, S_e, \rho_s$) are re-inserted into the 3D profile equations to find the point of steepest slope estimated from the projected profiles, which we denote as $R^p_{\rm st}$, as well as the prominence of the splashback feature. 

\begin{figure}
    \centering
    \includegraphics[width=\linewidth]{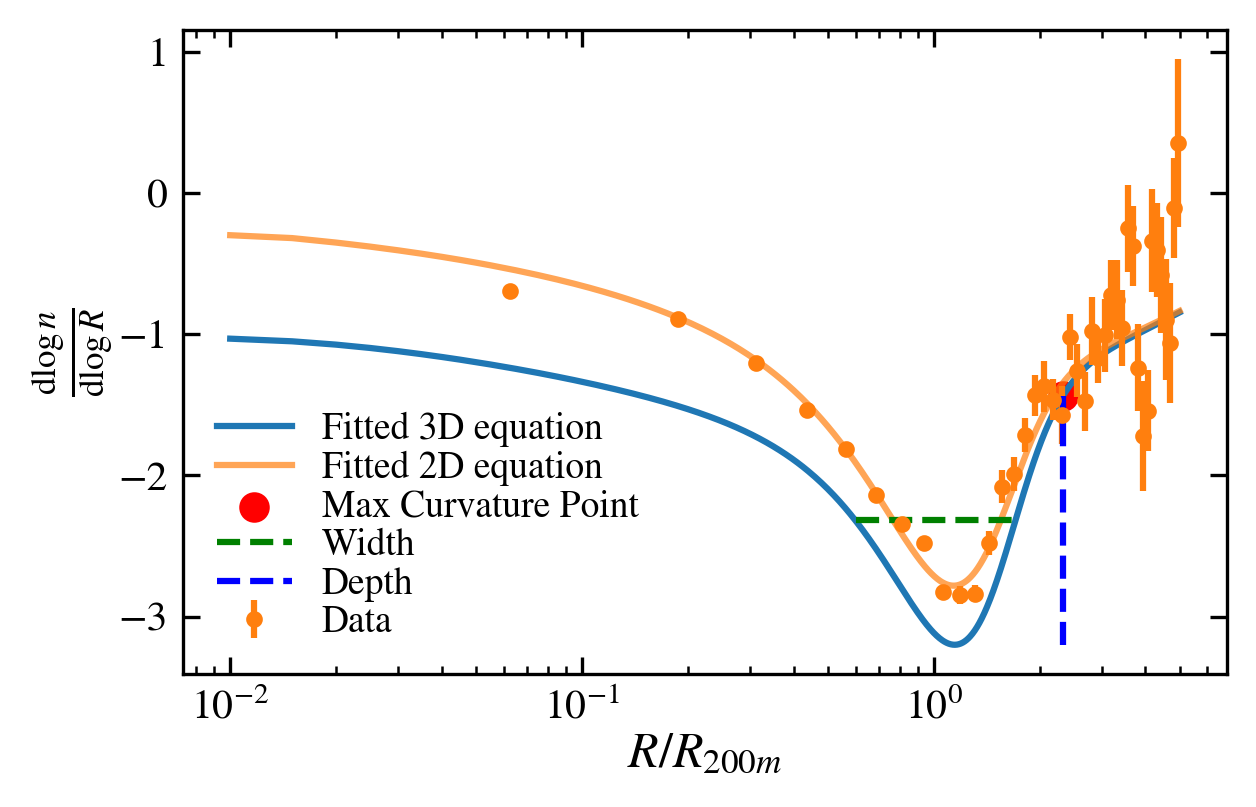}
    \caption{Example of how the projected derivative is fitted, using $d_{\rm max}=1000$ kpc, $M/\msun \in \left[ 10^{13.5},10^{14}\right)$. The orange points with error bars are one of the 2048 ``outer'' bootstraps, and the orange line is from fitting the projected derivative. Using those parameters in the 3D derivative gives the blue line. The point of maximum curvature is marked with a dot, and the corresponding depth and width shown.}
    \label{fig:example_fit}
\end{figure}

\begin{figure*}
    \centering
    \includegraphics[width=0.95\linewidth]{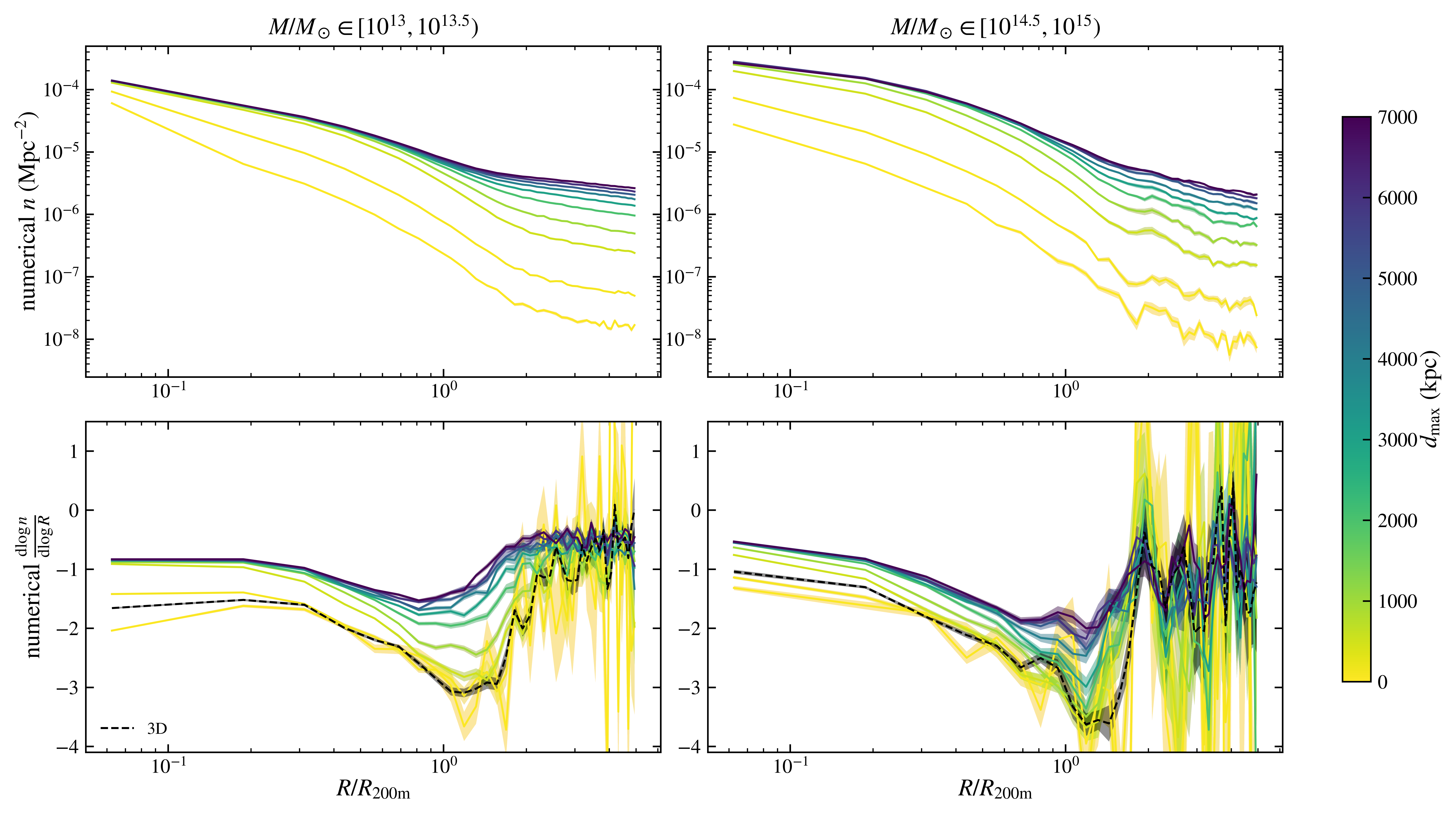}
    \caption{Examples of projected density profiles (top) and their 2D derivatives (bottom), from different projection lengths, for the lowest and highest mass bin. Each line is the mean profile of 32 inner bootstraps, and the shaded region is within $1\sigma$ of the mean. The derivative from the 3D profile is shown in dashed black on the bottom row. The 2D derivative at low projection lengths is closer in shape to the 3D derivative, but has greater uncertainty due to fewer galaxies counted.}
    \label{fig:all_profiles_and_derivs}
\end{figure*}

\begin{figure*}
    \centering
    \includegraphics[width=0.9\linewidth]{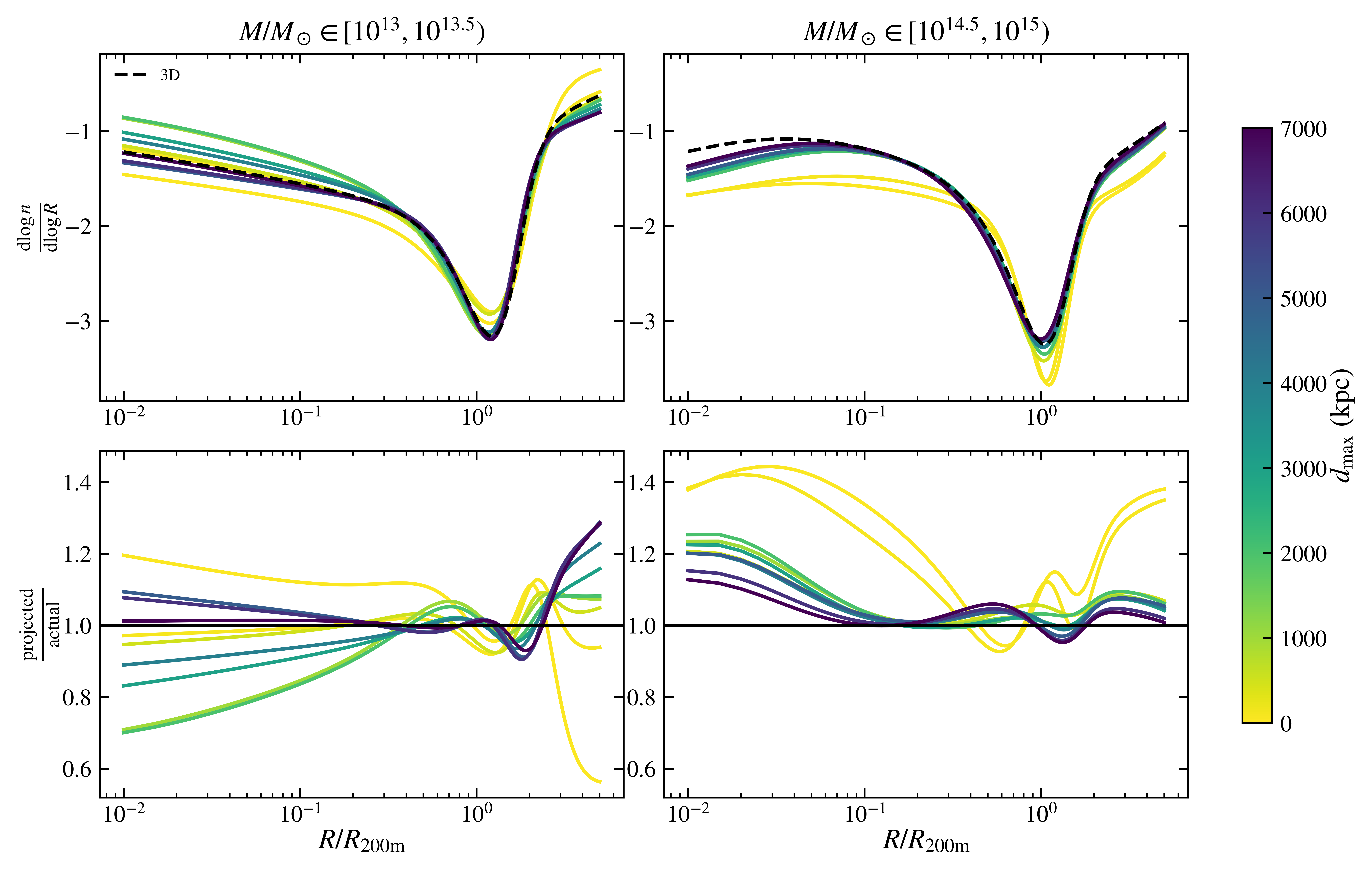}
    \caption{The 3D derivatives determined from parameters from the projected profile fits, compared to the 3D derivative from the true 3D profile (dashed black), for the largest and smallest mass bins, without bootstrapping. The top row shows the derivatives themselves, and the bottom row shows the fractional difference between the projected fits and the 3D fit. The recovered 3D profiles do not differ significantly from the true 3D profile, especially at larger projection lengths.}
    \label{fig:diffs_derivs}
\end{figure*}

\subsection{Prominence of splashback feature}
\label{sec:methods_feature}
To quantify the effects of projection in ``smearing'' out the splashback feature, we define an empirical measure of the prominence of the splashback feature using the depth and width of $\frac{\mathrm{d} \, \log n}{\mathrm{d} \, \log r}$. The point of maximum curvature of $\frac{\mathrm{d} \, \log n}{\mathrm{d} \, \log r}$ seems to correspond to where the number density begins to flatten out to the background density. Thus, the depth is measured as the difference in $\frac{\mathrm{d} \, \log n}{\mathrm{d} \, \log r}$ between this point of maximum curvature, and the point of $R_{\rm st}$.
The width is the width of $\frac{\mathrm{d} \, \log n}{\mathrm{d} \, \log r}$ halfway down the depth (Figure \ref{fig:example_fit}). Then, the ratio $\frac{\mathrm{depth}}{\mathrm{width}}$ can be used as a measure of splashback feature prominence, where a greater depth means the gradient of the density profile is the steepest, and a smaller width means the sharp change in density is most localized.

%% file: results.tex
\section{Results}
\label{sec:results}

\input{results_projection}

%% file: results_projection.tex
\label{sec:results_projection}

In this section, we explore the effects of the projection length $d_{\rm max}$ on the observed splashback feature.

\citet{Chang_2018} set $d_{\rm max} = 40 h^{-1}$ Mpc and found that changing $d_{\rm max}$ by 50\% does not significantly affect density profiles. 
Similarly, \citet{More_2016} used $d_{\rm max} = 40 h^{-1}$ Mpc as their default value, noting that the location of the splashback radius seems insensitive to this choice. This corresponds to a physical length of about 60 Mpc.
Although this is the maximum integrated projection depth, a galaxy's contribution to the density profile is weighted by its membership probability, so these studies do not include every galaxy within their projected cylinder as we do here.

Here, we investigate a range of projection lengths from 30 to 7000 kpc. We demonstrate that although smaller projection lengths ideally give an unbiased ``slice'' of the 3D density profile, larger projection lengths would include more galaxies, which could improve the estimate of the number density. 
By projecting all galaxies within the projection length, galaxies that are not part of the cluster are accounted for, but the impact of their inclusion is outweighed by increasing the prominence of the feature by having a larger number of galaxies. 

Using the methods described in the above sections, for each mass bin and projection length, we obtain 2048 bootstrapped profiles with error bars. We observe in Figure~\ref{fig:all_profiles_and_derivs} that the steepening in the derivative is still clearly identifiable, even in the projected profiles up to 7000 kpc, although this steepening does not necessarily bring the derivative below -3 as in the NFW profile. The smaller projection lengths result in a 2D profile and derivative closer in shape to the 3D case. However, since fewer galaxies are counted, there is greater error and less smoothness even in the stacked derivatives, which could adversely affect the fitting and calculation of $R_{\rm st}$. In fact, projection lengths below 30 kpc were not investigated since the fitting often did not converge due to noisiness of the profiles.
This is especially important when $N_{\rm halos}$ stacked is small, such as in the highest mass bin in Figure \ref{fig:all_profiles_and_derivs}. \citet{More_2016} stacks $4000-8000$ halos while \citet{Chang_2018} stacks $1000-3000$ halos to determine the profile. However, if our subsample of $N_{\rm halos}$ is smaller, for example if we use smaller mass bins to investigate the trend of $R_{\rm st}$ with halo mass, then either the number of halos in each bin must be large, or the projection length must be large, to reduce uncertainties in the profiles that adversely affect the fit.

\begin{figure}
    \centering
    \includegraphics[width=\linewidth]{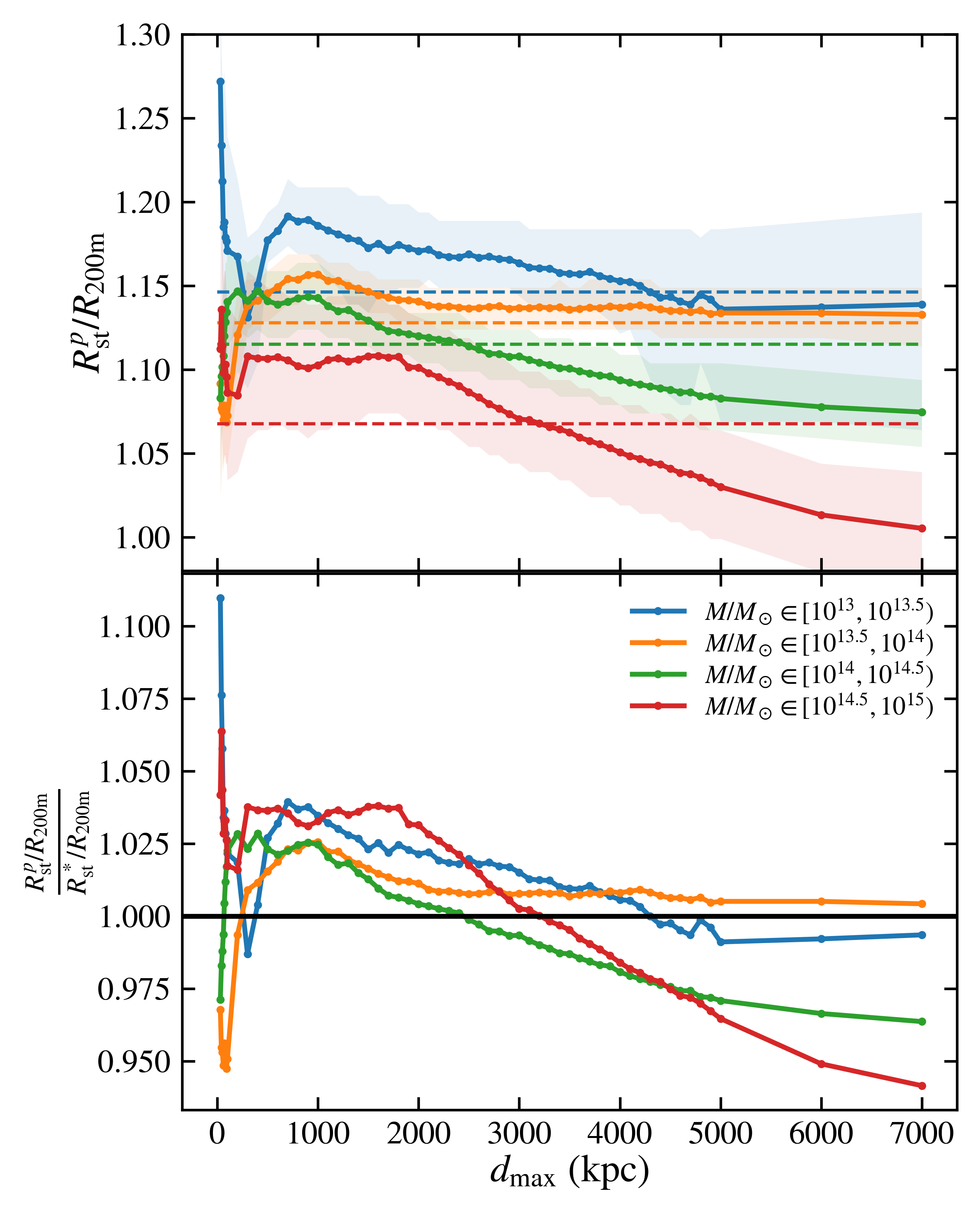}
    \caption{Top: Normalised $R^p_{\rm st}$ obtained using the projected profiles, against projection length. The shaded region is the 16th-84th percentile (of the 2048 bootstraps). The horizontal dotted lines represent normalised $R^*_{\rm st}$ obtained from the 3D profiles. Bottom: Fractional difference between $R^p_{\rm st}$ and $R^*_{\rm st}$, against projection length. $R^p_{\rm st}$ tracks $R^*_{\rm st}$ with a decreasing trend, especially for larger masses, that could be due to the change in shape of the feature biasing the fit.}
    \label{fig:rsp}
\end{figure}

\begin{figure}
    \centering
    \includegraphics[width=0.95\linewidth]{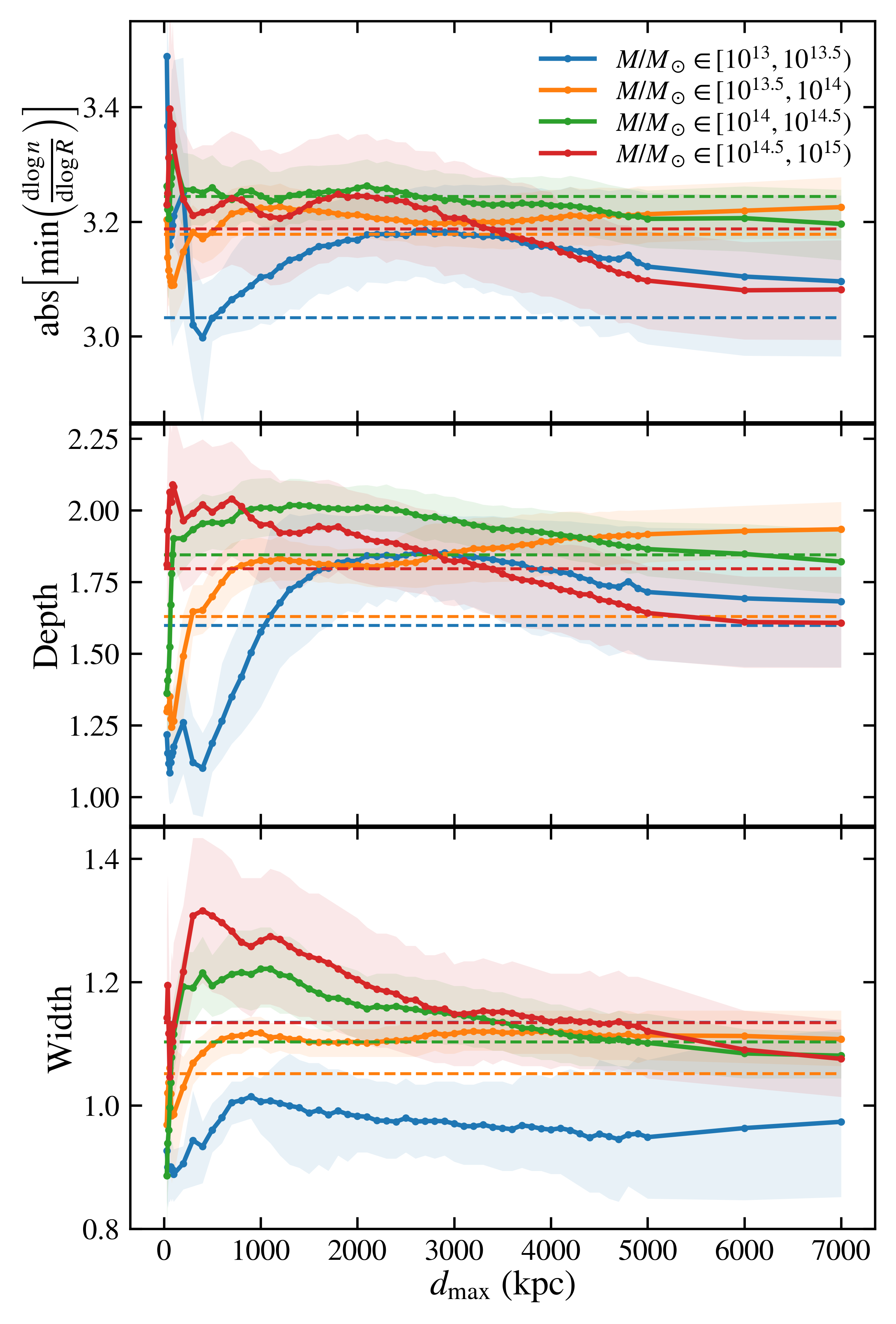}
    \caption{Measures of prominence of the splashback feature, against projection length, for the 3D derivatives obtained from the projected profiles. These measures do not change significantly with projection length. Top: absolute value of the minimum derivative. Middle: depth of derivative, as defined in section \ref{sec:methods_feature} Bottom: width of the derivative, as defined in section \ref{sec:methods_feature}.}
    \label{fig:depths_and_widths}
    \vspace{.2cm}
\end{figure}

Since we determine cluster membership by simply considering the cylinder of galaxies around the true halo centre, larger projection lengths may also reduce the impact of uncertainties in the line-of-sight distance between the galaxies and the halo centre. However, larger projection lengths means we end up observing the density drop for galaxies not in the plane of the halo centre, which, when projected, appear closer to the halo centre. This leads to the density drop of the 2D profile $R^{\rm 2D}_{\rm st}$ occurring at lower radii. This can be seen in the derivative as the minimum point shifts leftward with larger projection lengths. \citet{towler2024} projected the fitted 3D density profiles and found that the steepening in the 2D profiles occurs at radius $0.82 \pm 0.03$ times that of the 3D profiles, which is similar to what we observe here.

This effect is corrected once we use the fitted 2D parameters to calculate the inferred 3D profile and derivative, which end up being similar to the actual 3D profile. Figure \ref{fig:diffs_derivs} shows the fitted derivatives for the bottom and top mass bin, without bootstrapping, from both the true 3D profile and that inferred from the projected profiles. The fractional differences show that the discrepancy is indeed small, especially near the splashback feature.

\begin{figure}
    \centering
    \includegraphics[width=0.97\linewidth]{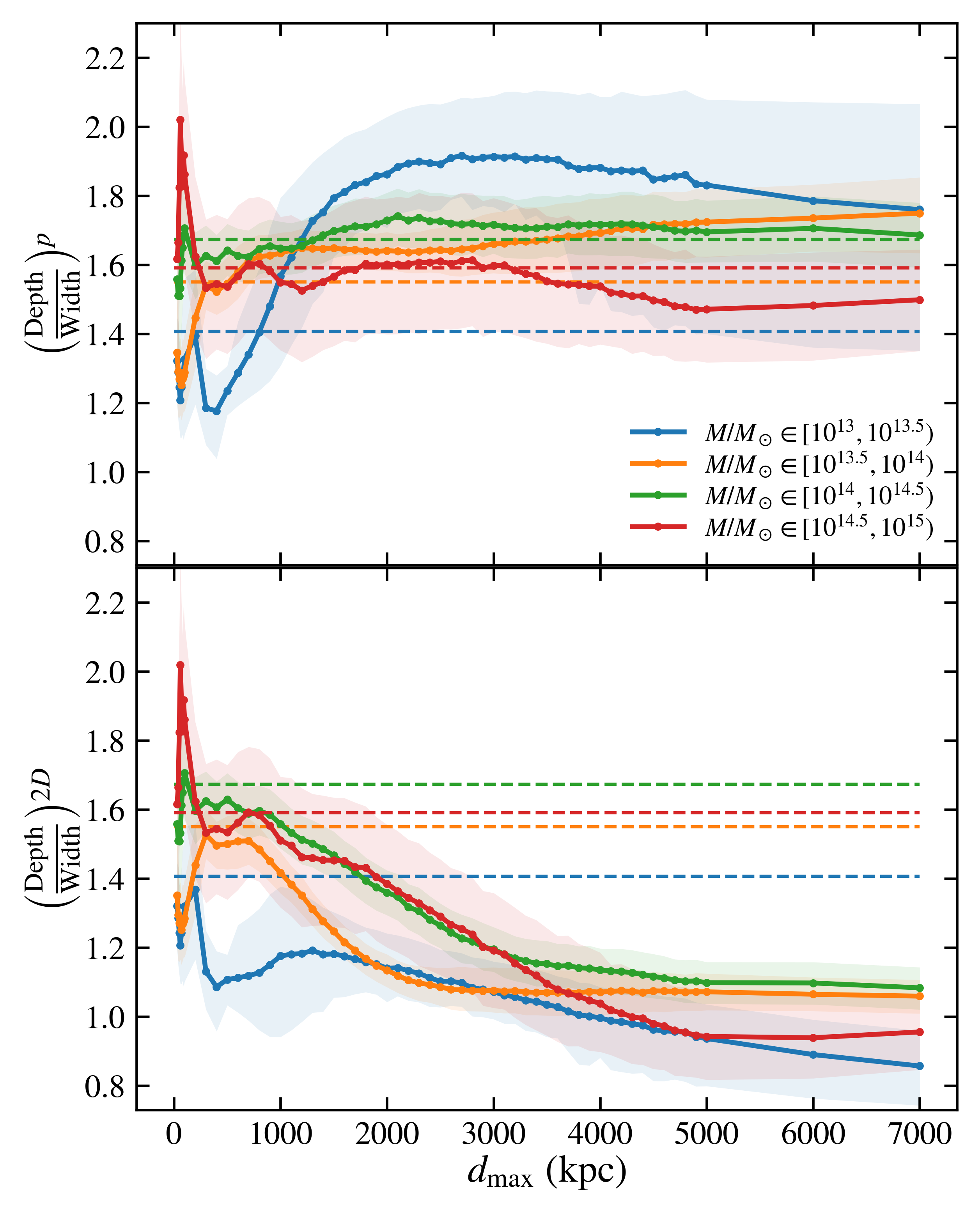}
    \caption{Ratios of depth and width against projection length. Top: ratios determined from the 3D derivatives obtained from the projected profiles. Bottom: ratios directly determined from 2D derivatives. The horizontal dotted lines represent the true 3D values. The splashback feature is less prominent at large projection lengths only in the 2D projected derivatives, but stays relatively constant with projection length in the inferred 3D derivatives.}
    \label{fig:ratios}
\end{figure}

Plotting the splashback radius obtained from the projected profiles $R^p_{\rm st}$ against the projection length (Figure \ref{fig:rsp}), we observe that $R^p_{\rm st}$ from the inferred 3D profile closely tracks the $R^*_{\rm st}$ from the true 3D profile for the lower mass bins. The discrepancy and error bars are largest below $\sim$1 Mpc, which aligns with the hypothesis that too-small projection lengths may lead to greater error in the number density profiles and thus the fitting procedure.

As the projection length increases, there is a decreasing trend in $R^p_{\rm st}$ derived from the projected profile, where $R^p_{\rm st}$ generally goes from overestimating $R^*_{\rm st}$ to underestimating it. In the 2D derivatives, as projection length increases, the splashback feature is shallower and occurs at a much smaller radius as seen in Figure \ref{fig:all_profiles_and_derivs}. Although this is mostly corrected for by using the fitted parameters in the 3D equation, the change in the prominence and shape of the feature itself could bias the fitting procedure, and lead to the observed change in $R^p _{\rm st}$ of up to 5-10\% that is purely a consequence of the projection.  The trend is more significant in the larger mass bins because the projection depth is a smaller fraction of the cluster radius.  The trend flattens in the lower mass bins as the projection depth envelops the cluster and not much is added as the projection depth increases. There is some overestimation of $R_{\rm st}$ at lower $d_{\rm max}$.
This is likely an artefact of the integration because the profile is not averaged over as large a projection depth.
As seen in Figure \ref{fig:diffs_derivs}, the shape of the profile is over-corrected when recovering the 3D profile from the 2D fits.

To quantify the prominence of the splashback feature itself, we investigate its depth and width as defined in Section \ref{sec:methods_feature}.
As shown previously in \citet{O_Neil_2021}, the prominence and shape of the splashback feature can alter the recovered value for $R_{\rm st}$.
Additionally, a wide and shallow feature may have a larger error.
The physical width of the splashback feature is predicted to increase with the velocity dispersion of dark matter particles \citep{Mohayaee2006} which in turn increases with the halo mass.
It is also predicted that the strength of the caustic can bias mass measurements from weak lensing \citep{Zhang2023}, especially in lower mass clusters and groups with mass around $M\sim10^{13.5}$.
Thus, it is important to establish the effect the projection has on smearing the splashback feature to disentangle this from the physical properties of the splashback feature.

We indeed observe in Figure~\ref{fig:depths_and_widths} that the width of the splashback feature is larger for larger mass bins. With increasing projection length, the minimum derivative, depth, and width of the splashback feature in the inferred 3D profile do not trend significantly, showing that projection to greater depths does not necessarily smear out the resulting splashback feature in the inferred 3D profiles, despite appearing to do so in the 2D profiles. 

This is confirmed by plotting the ratio of depth to width of the splashback feature against projection length. For the 2D derivative, as the projection length increases, the feature is increasingly ``smeared out'' and less prominent (Figure \ref{fig:all_profiles_and_derivs}), thus the ratio decreases (Figure \ref{fig:ratios}).  Using the fitted parameters in the 3D equation corrects for the smearing caused by projection, thus the ratio remains relatively consistent in projection length.  We suggest using a projection length of at least 2 Mpc, where it appears that the quantities measured here become stable even at lower mass clusters. However, since fitting is done on the 2D derivative, the parameters could be biased by the changing shape of the feature.

%% file: conclusions.tex
\section{Conclusions}
\label{sec:conclusions}

Observational studies of the splashback radius of galaxy clusters have shown some discrepancy with the expected values from simulations. The splashback radius can be approximated by fitting the number density profile of galaxies and finding the point of steepest slope $R_{\rm st}$. We study the effect of projection onto the plane of the sky and inclusion of galaxies within this projection on these profiles on the measurement of $R_{\rm st}$. Our conclusions are summarised as follows.

\begin{enumerate}
    \item In projected stacked profiles, the splashback feature can be noisy for small projection lengths due to the presence of fewer galaxies counted. As the projection length increases, the feature in the projected profiles shifts to a smaller radius and becomes shallower (Figure \ref{fig:all_profiles_and_derivs}).

    \item Fitting the projected number density equation \ref{eq:2D} with the projected profile results in a 3D profile and derivative similar to the true 3D profile. The resulting $R^p_{\rm st}$ is close to the true $R^*_{\rm st}$, but decreases with increasing projection length (Figure \ref{fig:rsp}). This is likely because longer projection lengths have shallower splashback features that bias the fit towards smaller $R^p_{\rm st}$.  The trend flattens out in lower mass bins more quickly than in larger mass bins due to $d_{\rm max}$ enclosing a larger fraction of the cluster.

    \item We quantify the prominence of the splashback feature using the depth and width of the derivative, where a higher ratio of depth to width is a more prominent feature, to confirm that the prominence decreases with projection length in the 2D profiles but remains constant with projection length in the 3D profiles (Figure \ref{fig:depths_and_widths} \& \ref{fig:ratios}).  The shape of the splashback feature appears to stabilize at a projection depth of roughly 2 Mpc.

\end{enumerate}

Although there remain several observational errors, such as galaxy cluster selection, halo centering, and cluster galaxy membership, we show here that projection effects are not a major source of error in reconstructing the shape of the density profiles at large radii. Halo mis-centring would affect the profile shape, likely decreasing the central density and biasing $R_{\rm st}$ to lower values, and this effect can be tested systematically using a similar method in future work.
The mis-centring of haloes is typically on the order of a few kpc \citep{Roche2024}, so we expect this effect to be secondary to projection.

Contamination of galaxies outside the projection length may also increase uncertainty, since observations have access to only line-of-sight velocities rather than exact positions as in simulations.
However, we do not expect this to significantly alter the results.
Our work shows that extending even well past the virial radius only moderately alters the results, and including a few more galaxies or missing a few galaxies that fall within this length is unlikely to significantly alter the density profile.
Additionally, weighting the contribution to the density profile by membership probability should mitigate this.

Despite the shape of the profile being somewhat altered at low projection lengths, larger projection lengths result in a recovered 3D profile that is very similar to the true 3D profile. It would, however, be prudent to remain somewhat cautious of the actual values of $R_{\rm st}$ if particularly precise measurements are required, given the decreasing trend in $R_{\rm st}$ with projection length.  It is likely that 2D measurements can be used to infer properties derived from the splashback feature, such as relations with the accretion history \citep{Shin2023} or inferences of galaxy cluster mass \citep{Zhang2023}.
With the depth and width of the splashback feature being measurable quantities, we may also be able to infer information about a halo's assembly history, peak height, and mass \citep{Yu2025}.

%% file: appendix_fit.tex
\section{Fit parameters}
\label{apx:fits}

Here we report our fit parameters for various projection depths in each mass bin for reference.

\begin{table}[H]
    \centering
    \begin{tabular}{c|cccc}
        \hline
        $d_{\rm max}$ (kpc) & 1000 & 3000 & 5000 & 3D \\
        \hline
        $r_s$     & 0.905 & 0.798 & 0.818 & 0.842 \\
        $r_t$     & 1.42 & 1.35 & 1.35 & 1.86 \\
        $\alpha$  & 0.0979 & 0.0896 & 0.0919 & 0.418 \\
        $\beta$   & 3.29 & 3.65 & 3.56 & 2.35 \\
        $\gamma$  & 7.88 & 7.86 & 8.07 & 9.27 \\
        $b_e$     & 1.26 & 1.27 & 1.26 & 0.383 \\
        $S_e$     & 1.6 & 1.41 & 1.48 & 2.02 \\
        $\rho_s/10^{-9}$  &2.51&2.72& 2.52 & 1.64\\
        \hline
    \end{tabular}
    \caption{Mean fitted parameters for mass bin $\left[10^{13},10^{13.5}\right)$ at different projection depths.}
\end{table}

\begin{table}[H]
    \centering
    \begin{tabular}{c|cccc}
        \hline
        $d_{\rm max}$ (kpc) & 1000 & 3000 & 5000 & 3D \\
        \hline
        $r_s$     & 0.936 & 0.929 & 0.929 & 0.960 \\
        $r_t$     & 1.52 & 1.52 & 1.52 & 1.55 \\
        $\alpha$  & 0.184 & 0.187 & 0.183 & 0.226 \\
        $\beta$   & 2.57 & 2.54 & 2.57 & 2.50 \\
        $\gamma$  & 7.74 & 7.76 & 7.91 & 8.33 \\
        $b_e$     & 1.29 & 1.30 & 1.36 & 1.17 \\
        $S_e$     & 1.47 & 1.41 & 1.35 & 1.65 \\
        $\rho_s/10^{-9}$  & 3.27 & 3.12& 3.06 & 3.05\\
        \hline
    \end{tabular}
    \caption{Mean fitted parameters for mass bin $\left[10^{13.5},10^{14}\right)$ at different projection depths.}
\end{table}

\begin{table}[H]
    \centering
    \begin{tabular}{c|cccc}
        \hline
        $d_{\rm max}$ (kpc) & 1000 & 3000 & 5000 & 3D \\
        \hline
        $r_s$     & 0.944 & 0.906 & 0.925 & 0.966 \\
        $r_t$     & 1.66 & 1.61 & 1.61 & 1.61 \\
        $\alpha$  & 0.248 & 0.255 & 0.258 & 0.247 \\
        $\beta$   & 2.13 & 2.22 & 2.21 & 2.29 \\
        $\gamma$  & 7.47 & 7.79 & 8.25 & 8.36 \\
        $b_e$     & 1.27 & 1.35 & 1.38 & 1.22 \\
        $S_e$     & 1.29 & 1.32 & 1.39 & 2.02 \\
        $\rho_s/10^{-9}$  &3.65&3.82& 3.70 & 3.39\\
        \hline
    \end{tabular}
    \caption{Mean fitted parameters for mass bin $\left[10^{14},10^{14.5}\right)$ at different projection depths.}
\end{table}

\begin{table}[H]
    \centering
    \begin{tabular}{c|cccc}
        \hline
        $d_{\rm max}$ (kpc) & 1000 & 3000 & 5000 & 3D \\
        \hline
        $r_s$     &1.09&1.05&0.974&1.07\\
        $r_t$     &1.89&1.70&1.75&1.83\\
        $\alpha$  &0.306&0.301&0.336&0.295\\
        $\beta$   &1.67&1.87&1.75&1.84\\
        $\gamma$  &7.60&7.85&7.75&8.84\\
        $b_e$     &1.14&1.30&1.42&1.47\\
        $S_e$     &1.33&1.43&1.42&1.47\\
        $\rho_s/10^{-9}$  &3.89&3.93&4.67&3.55 \\
        \hline
    \end{tabular}
    \caption{Mean fitted parameters for mass bin $\left[10^{14.5},10^{15}\right)$ at different projection depths.}
\end{table}

%% file: main.bbl
\begin{thebibliography}{50}
\expandafter\ifx\csname natexlab\endcsname\relax\def\natexlab#1{#1}\fi

\bibitem[{{Adhikari}, {Dalal} \& {Chamberlain}(2014){Adhikari}, {Dalal}, \& {Chamberlain}}]{Adhikari2014}
{Adhikari} S., {Dalal} N., {Chamberlain} R.~T., 2014, \jcap, 2014, 019

\bibitem[{{Adhikari} {et~al}\mbox{.}(2021){Adhikari}, {Shin}, {Jain}, {Hilton}, {Baxter}, {Chang}, {Wechsler}, {Battaglia}, {Bond}, {Bocquet}, {Choi}, {DeRose}, {Devlin}, {Dunkley}, {Evrard}, {Ferraro}, {Hill}, {Hughes}, {Gallardo}, {Lokken}, {MacInnis}, {Madhavacheril}, {McMahon}, {Nati}, {Newburgh}, {Niemack}, {Page}, {Palmese}, {Partridge}, {Rozo}, {Rykoff}, {Salatino}, {Schillaci}, {Sehgal}, {Sif{\'o}n}, {To}, {Wollack}, {Wu}, {Xu}, {Aguena}, {Allam}, {Amon}, {Annis}, {Avila}, {Bacon}, {Bertin}, {Bhargava}, {Brooks}, {Burke}, {Rosell}, {Kind}, {Carretero}, {Castander}, {Choi}, {Costanzi}, {da Costa}, {Vicente}, {Desai}, {Diehl}, {Doel}, {Everett}, {Ferrero}, {Fert{\'e}}, {Flaugher}, {Fosalba}, {Frieman}, {Garc{\'\i}a-Bellido}, {Gaztanaga}, {Gruen}, {Gruendl}, {Gschwend}, {Gutierrez}, {Hartley}, {Hinton}, {Hollowood}, {Honscheid}, {James}, {Jeltema}, {Kuehn}, {Kuropatkin}, {Lahav}, {Lima}, {Maia}, {Marshall}, {Martini}, {Melchior}, {Menanteau}, {Miquel}, {Morgan}, {L.~C. Ogando}, {Paz-Chinch{\'o}n},
  {Malag{\'o}n}, {Sanchez}, {Santiago}, {Scarpine}, {Serrano}, {Sevilla-Noarbe}, {Smith}, {Soares-Santos}, {Suchyta}, {E.~C. Swanson}, {Varga}, {Wilkinson}, {Zhang}, {Austermann}, {Beall}, {Becker}, {Denison}, {Duff}, {Hilton}, {Hubmayr}, {Ullom}, {Lanen}, {Vale}, {Vale}, \& {Vale}}]{Adhikari2021}
{Adhikari} S. {et~al.}, 2021, \apj, 923, 37

\bibitem[{{Baxter} {et~al}\mbox{.}(2017){Baxter}, {Chang}, {Jain}, {Adhikari}, {Dalal}, {Kravtsov}, {More}, {Rozo}, {Rykoff}, \& {Sheth}}]{baxter2017}
{Baxter} E. {et~al.}, 2017, \apj, 841, 18

\bibitem[{{Busch} \& {White}(2017)}]{BuschWhite}
{Busch} P., {White} S. D.~M., 2017, \mnras, 470, 4767

\bibitem[{Chang {et~al}\mbox{.}(2018)Chang, Baxter, Jain, Sánchez, Adhikari, Varga, Fang, Rozo, Rykoff, Kravtsov, Gruen, Hartley, Huff, Jarvis, Kim, Prat, MacCrann, McClintock, Palmese, Rapetti, Rollins, Samuroff, Sheldon, Troxel, Wechsler, Zhang, Zuntz, Abbott, Abdalla, Allam, Annis, Bechtol, Benoit-Lévy, Bernstein, Brooks, Buckley-Geer, Rosell, Kind, Carretero, D’Andrea, Costa, Davis, Desai, Diehl, Dietrich, Drlica-Wagner, Eifler, Flaugher, Fosalba, Frieman, García-Bellido, Gaztanaga, Gerdes, Gruendl, Gschwend, Gutierrez, Honscheid, James, Jeltema, Krause, Kuehn, Lahav, Lima, March, Marshall, Martini, Melchior, Menanteau, Miquel, Mohr, Nord, Ogando, Plazas, Sanchez, Scarpine, Schindler, Schubnell, Sevilla-Noarbe, Smith, Smith, Soares-Santos, Sobreira, Suchyta, Swanson, Tarle, \& Weller}]{Chang_2018}
Chang C. {et~al.}, 2018, The Astrophysical Journal, 864, 83

\bibitem[{{Cooper} {et~al}\mbox{.}(2006){Cooper}, {Newman}, {Croton}, {Weiner}, {Willmer}, {Gerke}, {Madgwick}, {Faber}, {Davis}, {Coil}, {Finkbeiner}, {Guhathakurta}, \& {Koo}}]{Cooper2006}
{Cooper} M.~C. {et~al.}, 2006, \mnras, 370, 198

\bibitem[{{Diemer} \& {Kravtsov}(2014)}]{Diemer2014}
{Diemer} B., {Kravtsov} A.~V., 2014, \apj, 789, 18

\bibitem[{Diemer {et~al}\mbox{.}(2017)Diemer, Mansfield, Kravtsov, \& More}]{Diemer_2017}
Diemer B., Mansfield P., Kravtsov A.~V., More S., 2017, The Astrophysical Journal, 843, 140

\bibitem[{{Diemer}, {More} \& {Kravtsov}(2013){Diemer}, {More}, \& {Kravtsov}}]{Diemer2013}
{Diemer} B., {More} S., {Kravtsov} A.~V., 2013, \apj, 766, 25

\bibitem[{{Dolag} {et~al}\mbox{.}(2009){Dolag}, {Borgani}, {Murante}, \& {Springel}}]{Dolag2009}
{Dolag} K., {Borgani} S., {Murante} G., {Springel} V., 2009, \mnras, 399, 497

\bibitem[{{Donnari} {et~al}\mbox{.}(2021){Donnari}, {Pillepich}, {Joshi}, {Nelson}, {Genel}, {Marinacci}, {Rodriguez-Gomez}, {Pakmor}, {Torrey}, {Vogelsberger}, \& {Hernquist}}]{Donnari2021}
{Donnari} M. {et~al.}, 2021, \mnras, 500, 4004

\bibitem[{{Dressler}(1980)}]{Dressler1980}
{Dressler} A., 1980, \apj, 236, 351

\bibitem[{Einasto(1965)}]{einasto1965trudy}
Einasto J., 1965, Alma-Ata, 5, 87

\bibitem[{{Gunn} \& {Gott}(1972)}]{gunn1972}
{Gunn} J.~E., {Gott}, III J.~R., 1972, \apj, 176, 1

\bibitem[{{Harris} {et~al}\mbox{.}(2020){Harris}, {Millman}, {van der Walt}, {Gommers}, {Virtanen}, {Cournapeau}, {Wieser}, {Taylor}, {Berg}, {Smith}, {Kern}, {Picus}, {Hoyer}, {van Kerkwijk}, {Brett}, {Haldane}, {del R{\'\i}o}, {Wiebe}, {Peterson}, {G{\'e}rard-Marchant}, {Sheppard}, {Reddy}, {Weckesser}, {Abbasi}, {Gohlke}, \& {Oliphant}}]{Harris2020}
{Harris} C.~R. {et~al.}, 2020, \nat, 585, 357

\bibitem[{{Hunter}(2007)}]{Hunter2007}
{Hunter} J.~D., 2007, Computing in Science and Engineering, 9, 90

\bibitem[{{Klein} {et~al}\mbox{.}(2019){Klein}, {Grandis}, {Mohr}, {Paulus}, {Abbott}, {Annis}, {Avila}, {Bertin}, {Brooks}, {Buckley-Geer}, {Rosell}, {Kind}, {Carretero}, {Castander}, {Cunha}, {D'Andrea}, {da Costa}, {De Vicente}, {Desai}, {Diehl}, {Dietrich}, {Doel}, {Evrard}, {Flaugher}, {Fosalba}, {Frieman}, {Garc{\'\i}a-Bellido}, {Gaztanaga}, {Giles}, {Gruen}, {Gruendl}, {Gschwend}, {Gutierrez}, {Hartley}, {Hollowood}, {Honscheid}, {Hoyle}, {James}, {Jeltema}, {Kuehn}, {Kuropatkin}, {Lima}, {Maia}, {March}, {Marshall}, {Menanteau}, {Miquel}, {Ogando}, {Plazas}, {Romer}, {Roodman}, {Sanchez}, {Scarpine}, {Schindler}, {Serrano}, {Sevilla-Noarbe}, {Smith}, {Smith}, {Soares-Santos}, {Sobreira}, {Suchyta}, {Swanson}, {Tarle}, {Thomas}, {Vikram}, \& {DES Collaboration}}]{Klein2019}
{Klein} M. {et~al.}, 2019, \mnras, 488, 739

\bibitem[{{Koester} {et~al}\mbox{.}(2007){Koester}, {McKay}, {Annis}, {Wechsler}, {Evrard}, {Bleem}, {Becker}, {Johnston}, {Sheldon}, {Nichol}, {Miller}, {Scranton}, {Bahcall}, {Barentine}, {Brewington}, {Brinkmann}, {Harvanek}, {Kleinman}, {Krzesinski}, {Long}, {Nitta}, {Schneider}, {Sneddin}, {Voges}, \& {York}}]{Koester2007}
{Koester} B.~P. {et~al.}, 2007, \apj, 660, 239

\bibitem[{{Lim} {et~al}\mbox{.}(2021){Lim}, {Barnes}, {Vogelsberger}, {Mo}, {Nelson}, {Pillepich}, {Dolag}, \& {Marinacci}}]{Lim2021}
{Lim} S.~H., {Barnes} D., {Vogelsberger} M., {Mo} H.~J., {Nelson} D., {Pillepich} A., {Dolag} K., {Marinacci} F., 2021, \mnras, 504, 5131

\bibitem[{Mansfield, Kravtsov \& Diemer(2017)Mansfield, Kravtsov, \& Diemer}]{Mansfield_2017}
Mansfield P., Kravtsov A.~V., Diemer B., 2017, The Astrophysical Journal, 841, 34

\bibitem[{{Mohayaee} \& {Shandarin}(2006)}]{Mohayaee2006}
{Mohayaee} R., {Shandarin} S.~F., 2006, \mnras, 366, 1217

\bibitem[{{More}, {Diemer} \& {Kravtsov}(2015){More}, {Diemer}, \& {Kravtsov}}]{More2015}
{More} S., {Diemer} B., {Kravtsov} A.~V., 2015, \apj, 810, 36

\bibitem[{More, Diemer \& Kravtsov(2015)More, Diemer, \& Kravtsov}]{More_2015}
More S., Diemer B., Kravtsov A.~V., 2015, The Astrophysical Journal, 810, 36

\bibitem[{More {et~al}\mbox{.}(2016)More, Miyatake, Takada, Diemer, Kravtsov, Dalal, More, Murata, Mandelbaum, Rozo, Rykoff, Oguri, \& Spergel}]{More_2016}
More S. {et~al.}, 2016, The Astrophysical Journal, 825, 39

\bibitem[{{Murata} {et~al}\mbox{.}(2020){Murata}, {Sunayama}, {Oguri}, {More}, {Nishizawa}, {Nishimichi}, \& {Osato}}]{Murata2020}
{Murata} R., {Sunayama} T., {Oguri} M., {More} S., {Nishizawa} A.~J., {Nishimichi} T., {Osato} K., 2020, arxiv, 32

\bibitem[{{Nelson} {et~al}\mbox{.}(2019){Nelson}, {Springel}, {Pillepich}, {Rodriguez-Gomez}, {Torrey}, {Genel}, {Vogelsberger}, {Pakmor}, {Marinacci}, {Weinberger}, {Kelley}, {Lovell}, {Diemer}, \& {Hernquist}}]{Nelson2019}
{Nelson} D. {et~al.}, 2019, Computational Astrophysics and Cosmology, 6, 2

\bibitem[{Nelson {et~al}\mbox{.}(2021)Nelson, Springel, Pillepich, Rodriguez-Gomez, Torrey, Genel, Vogelsberger, Pakmor, Marinacci, Weinberger, Kelley, Lovell, Diemer, \& Hernquist}]{nelson2021illustristngsimulationspublicdata}
Nelson D. {et~al.}, 2021, The illustristng simulations: Public data release

\bibitem[{{Nishizawa} {et~al}\mbox{.}(2018){Nishizawa}, {Oguri}, {Oogi}, {More}, {Nishimichi}, {Nagashima}, {Lin}, {Mandelbaum}, {Takada}, {Bahcall}, {Coupon}, {Huang}, {Jian}, {Komiyama}, {Leauthaud}, {Lin}, {Miyatake}, {Miyazaki}, \& {Tanaka}}]{Nishizawa2018}
{Nishizawa} A.~J. {et~al.}, 2018, \pasj, 70, S24

\bibitem[{O'Neil {et~al}\mbox{.}(2021)O'Neil, Barnes, Vogelsberger, \& Diemer}]{O_Neil_2021}
O'Neil S., Barnes D.~J., Vogelsberger M., Diemer B., 2021, Monthly Notices of the Royal Astronomical Society, 504, 4649–4666

\bibitem[{{O'Neil} {et~al}\mbox{.}(2022){O'Neil}, {Borrow}, {Vogelsberger}, \& {Diemer}}]{O'Neil2022}
{O'Neil} S., {Borrow} J., {Vogelsberger} M., {Diemer} B., 2022, \mnras, 513, 835

\bibitem[{{O'Neil} {et~al}\mbox{.}(2024){O'Neil}, {Borrow}, {Vogelsberger}, {Zhao}, \& {Wang}}]{O'Neil2024}
{O'Neil} S., {Borrow} J., {Vogelsberger} M., {Zhao} H., {Wang} B., 2024, \mnras, 530, 3310

\bibitem[{{Pillepich} {et~al}\mbox{.}(2018){Pillepich}, {Nelson}, {Hernquist}, {Springel}, {Pakmor}, {Torrey}, {Weinberger}, {Genel}, {Naiman}, {Marinacci}, \& {Vogelsberger}}]{firstresultsTNG}
{Pillepich} A. {et~al.}, 2018, \mnras, 475, 648

\bibitem[{{Planck Collaboration} {et~al}\mbox{.}(2016){Planck Collaboration}, {Ade}, {Aghanim}, {Arnaud}, {Ashdown}, {Aumont}, {Baccigalupi}, {Banday}, {Barreiro}, {Bartlett}, {Bartolo}, {Battaner}, {Battye}, {Benabed}, {Beno{\^\i}t}, {Benoit-L{\'e}vy}, {Bernard}, {Bersanelli}, {Bielewicz}, {Bock}, {Bonaldi}, {Bonavera}, {Bond}, {Borrill}, {Bouchet}, {Boulanger}, {Bucher}, {Burigana}, {Butler}, {Calabrese}, {Cardoso}, {Catalano}, {Challinor}, {Chamballu}, {Chary}, {Chiang}, {Chluba}, {Christensen}, {Church}, {Clements}, {Colombi}, {Colombo}, {Combet}, {Coulais}, {Crill}, {Curto}, {Cuttaia}, {Danese}, {Davies}, {Davis}, {de Bernardis}, {de Rosa}, {de Zotti}, {Delabrouille}, {D{\'e}sert}, {Di Valentino}, {Dickinson}, {Diego}, {Dolag}, {Dole}, {Donzelli}, {Dor{\'e}}, {Douspis}, {Ducout}, {Dunkley}, {Dupac}, {Efstathiou}, {Elsner}, {En{\ss}lin}, {Eriksen}, {Farhang}, {Fergusson}, {Finelli}, {Forni}, {Frailis}, {Fraisse}, {Franceschi}, {Frejsel}, {Galeotta}, {Galli}, {Ganga}, {Gauthier}, {Gerbino}, {Ghosh},
  {Giard}, {Giraud-H{\'e}raud}, {Giusarma}, {Gjerl{\o}w}, {Gonz{\'a}lez-Nuevo}, {G{\'o}rski}, {Gratton}, {Gregorio}, {Gruppuso}, {Gudmundsson}, {Hamann}, {Hansen}, {Hanson}, {Harrison}, {Helou}, {Henrot-Versill{\'e}}, {Hern{\'a}ndez-Monteagudo}, {Herranz}, {Hildebrandt}, {Hivon}, {Hobson}, {Holmes}, {Hornstrup}, {Hovest}, {Huang}, {Huffenberger}, {Hurier}, {Jaffe}, {Jaffe}, {Jones}, {Juvela}, {Keih{\"a}nen}, {Keskitalo}, {Kisner}, {Kneissl}, {Knoche}, {Knox}, {Kunz}, {Kurki-Suonio}, {Lagache}, {L{\"a}hteenm{\"a}ki}, {Lamarre}, {Lasenby}, {Lattanzi}, {Lawrence}, {Leahy}, {Leonardi}, {Lesgourgues}, {Levrier}, {Lewis}, {Liguori}, {Lilje}, {Linden-V{\o}rnle}, {L{\'o}pez-Caniego}, {Lubin}, {Mac{\'\i}as-P{\'e}rez}, {Maggio}, {Maino}, {Mandolesi}, {Mangilli}, {Marchini}, {Maris}, {Martin}, {Martinelli}, {Mart{\'\i}nez-Gonz{\'a}lez}, {Masi}, {Matarrese}, {McGehee}, {Meinhold}, {Melchiorri}, {Melin}, {Mendes}, {Mennella}, {Migliaccio}, {Millea}, {Mitra}, {Miville-Desch{\^e}nes}, {Moneti}, {Montier}, {Morgante},
  {Mortlock}, {Moss}, {Munshi}, {Murphy}, {Naselsky}, {Nati}, {Natoli}, {Netterfield}, {N{\o}rgaard-Nielsen}, {Noviello}, {Novikov}, {Novikov}, {Oxborrow}, {Paci}, {Pagano}, {Pajot}, {Paladini}, {Paoletti}, {Partridge}, {Pasian}, {Patanchon}, {Pearson}, {Perdereau}, {Perotto}, {Perrotta}, {Pettorino}, {Piacentini}, {Piat}, {Pierpaoli}, {Pietrobon}, {Plaszczynski}, {Pointecouteau}, {Polenta}, {Popa}, {Pratt}, \& {Pr{\'e}zeau}}]{Planck2016}
{Planck Collaboration} {et~al.}, 2016, \aap, 594, A13

\bibitem[{{Roche} {et~al}\mbox{.}(2024){Roche}, {McDonald}, {Borrow}, {Vogelsberger}, {Shen}, {Springel}, {Hernquist}, {Pakmor}, {Bose}, \& {Kannan}}]{Roche2024}
{Roche} C. {et~al.}, 2024, The Open Journal of Astrophysics, 7, 65

\bibitem[{{Rozo} {et~al}\mbox{.}(2007){Rozo}, {Wechsler}, {Koester}, {Evrard}, \& {McKay}}]{Rozo2007}
{Rozo} E., {Wechsler} R.~H., {Koester} B.~P., {Evrard} A.~E., {McKay} T.~A., 2007, arXiv e-prints, astro

\bibitem[{{Rykoff} {et~al}\mbox{.}(2014){Rykoff}, {Rozo}, {Busha}, {Cunha}, {Finoguenov}, {Evrard}, {Hao}, {Koester}, {Leauthaud}, {Nord}, {Pierre}, {Reddick}, {Sadibekova}, {Sheldon}, \& {Wechsler}}]{Rykoff2014}
{Rykoff} E.~S. {et~al.}, 2014, \apj, 785, 104

\bibitem[{{Shin} {et~al}\mbox{.}(2019){Shin}, {Adhikari}, {Baxter}, {Chang}, {Battaglia}, {Bleem}, {Bocquet}, {DeRose}, {Gruen}, {Hilton}, {Kravtsov}, {McClintock}, {Rozo}, {Rykoff}, {Varga}, {Wechsler}, {Wu}, {Zhang}, {Aiola}, {Allam}, {Bechtol}, {Benson}, {Bertin}, {Bond}, {Brodwin}, {Brooks}, {Buckley-Geer}, {Burke}, {Carlstrom}, {Carrasco Kind}, {Carretero}, {Castander}, {Choi}, {Cunha}, {Crawford}, {da Costa}, {De Vicente}, {Desai}, {Devlin}, {Dietrich}, {Doel}, {Dunkley}, {Eifler}, {Evrard}, {Flaugher}, P., {Gallardo}, {Garc\'{i}a-Bellido}, {Gaztanaga}, {Gerdes}, {Gralla}, {Gruendl}, {Gschwend}, {Gitierrez}, {Hartley}, {Hill}, {Hollowood}, {Hoyle}, {Huffenberger}, {Hughes}, {James}, {Jeltema}, {Kim}, {Krause}, {Kuehn}, {Lahav}, {Lima}, {Madhavacheril}, {Maia}, {Marshall}, {Maurin}, {McMahon}, {Menanteau}, {Miller}, {Miquel}, {Mohr}, {Naess}, {Nati}, {Newburgh}, {Niemack}, {Ogando}, {Partridge}, {Patil}, {Plazas}, {Rapetti}, {Reichardt}, {Romer}, {Sanchez}, {Scarpine}, {Schindler}, {Serrano}, {Smith},
  {Smith}, {Soares-Santos}, {Sobreira}, {Staggs}, {Stark}, {Stein}, E., {Swanson}, {Tarle}, {Thomas}, {van Engelen}, {Wollack}, \& {Xu}}]{Shin2019}
{Shin} T. {et~al.}, 2019, \mnras, 487, 2900

\bibitem[{{Shin} {et~al}\mbox{.}(2021){Shin}, {Jain}, {Adhikari}, {Baxter}, {Chang}, {Pandey}, {Salcedo}, {Weinberg}, {Amsellem}, {Battaglia}, {Belyakov}, {Dacunha}, {Goldstein}, {Kravtsov}, {Varga}, {Abbott}, {Aguena}, {Alarcon}, {Allam}, {Amon}, {Andrade-Oliveira}, {Annis}, {Bacon}, {Bechtol}, {Becker}, {Bernstein}, {Bertin}, {Bocquet}, {Bond}, {Brooks}, {Buckley-Geer}, {Burke}, {Campos}, {Rosell}, {Kind}, {Carretero}, {Chen}, {Choi}, {Costanzi}, {da Costa}, {DeRose}, {Desai}, {De Vicente}, {Devlin}, {Diehl}, {Dietrich}, {Dodelson}, {Doel}, {Doux}, {Drlica-Wagner}, {Eckert}, {Elvin-Poole}, {Everett}, {Ferraro}, {Ferrero}, {Fert{\'e}}, {Flaugher}, {Frieman}, {Gallardo}, {Gatti}, {Gaztanaga}, {Gerdes}, {Gruen}, {Gruendl}, {Gutierrez}, {Harrison}, {Hartley}, {Hill}, {Hilton}, {Hinton}, {Hollowood}, {Hughes}, {James}, {Jarvis}, {Jeltema}, {Koopman}, {Krause}, {Kuehn}, {Kuropatkin}, {Lahav}, {Lima}, {Lokken}, {MacCrann}, {Madhavacheril}, {Maia}, {McCullough}, {McMahon}, {Melchior}, {Menanteau}, {Miquel}, {Mohr},
  {Moodley}, {Morgan}, {Myles}, {Nati}, {Navarro-Alsina}, {Niemack}, {Ogando}, {Page}, {Palmese}, {Partridge}, {Paz-Chinch{\'o}n}, {Pereira}, {Pieres}, {Malag{\'o}n}, {Prat}, {Raveri}, {Rodriguez-Monroy}, {Rollins}, {Romer}, {Rykoff}, {Salatino}, {S{\'a}nchez}, {Sanchez}, {Santiago}, {Scarpine}, {Schillaci}, {Secco}, {Serrano}, {Sevilla-Noarbe}, {Sheldon}, {Sherwin}, {Sif{\'o}n}, {Smith}, {Soares-Santos}, {Staggs}, {Suchyta}, {Swanson}, {Tarle}, {Thomas}, {To}, {Troxel}, {Tutusaus}, {Vavagiakis}, {Weller}, {Wollack}, {Yanny}, {Yin}, \& {Zhang}}]{Shin2021}
{Shin} T. {et~al.}, 2021, \mnras, 507, 5758

\bibitem[{{Shin} \& {Diemer}(2023)}]{Shin2023}
{Shin} T.-h., {Diemer} B., 2023, \mnras, 521, 5570

\bibitem[{{Smith} {et~al}\mbox{.}(2005){Smith}, {Kneib}, {Smail}, {Mazzotta}, {Ebeling}, \& {Czoske}}]{Smith2005}
{Smith} G.~P., {Kneib} J.-P., {Smail} I., {Mazzotta} P., {Ebeling} H., {Czoske} O., 2005, \mnras, 359, 417

\bibitem[{Springel(2010)}]{Springel_2010}
Springel V., 2010, Monthly Notices of the Royal Astronomical Society, 401, 791–851

\bibitem[{{Springel} {et~al}\mbox{.}(2001){Springel}, {White}, {Tormen}, \& {Kauffmann}}]{Springel2001}
{Springel} V., {White} S.~D.~M., {Tormen} G., {Kauffmann} G., 2001, \mnras, 328, 726

\bibitem[{{Sunayama} \& {More}(2019)}]{Sunayama2019}
{Sunayama} T., {More} S., 2019, \mnras, 490, 4945

\bibitem[{Towler {et~al}\mbox{.}(2024)Towler, Kay, Schaye, Kugel, Schaller, Braspenning, Elbers, Frenk, Kwan, Salcido, van Daalen, Vandenbroucke, \& Altamura}]{towler2024}
Towler I. {et~al.}, 2024, Inferring the dark matter splashback radius from cluster gas and observable profiles in the flamingo simulations

\bibitem[{Van~Rossum \& Drake~Jr(1995)}]{vanRossum1995}
Van~Rossum G., Drake~Jr F.~L., 1995, Python reference manual. Centrum voor Wiskunde en Informatica Amsterdam

\bibitem[{{Virtanen} {et~al}\mbox{.}(2020){Virtanen}, {Gommers}, {Oliphant}, {Haberland}, {Reddy}, {Cournapeau}, {Burovski}, {Peterson}, {Weckesser}, {Bright}, {van der Walt}, {Brett}, {Wilson}, {Millman}, {Mayorov}, {Nelson}, {Jones}, {Kern}, {Larson}, {Carey}, {Polat}, {Feng}, {Moore}, {VanderPlas}, {Laxalde}, {Perktold}, {Cimrman}, {Henriksen}, {Quintero}, {Harris}, {Archibald}, {Ribeiro}, {Pedregosa}, {van Mulbregt}, \& {SciPy 1. 0 Contributors}}]{Virtanen2020}
{Virtanen} P. {et~al.}, 2020, Nature Methods, 17, 261

\bibitem[{{Yu} {et~al}\mbox{.}(in prep){Yu}, {O'Neil}, {Shen}, {Vogelsberger}, {Wu}, \& {Wu}}]{Yu2025}
{Yu} Q.~S., {O'Neil} S., {Shen} X., {Vogelsberger} M., {Wu} M., {Wu} Z., in prep

\bibitem[{{Zhang} {et~al}\mbox{.}(2023){Zhang}, {Adhikari}, {Costanzi}, {Frieman}, {Annis}, \& {Chang}}]{Zhang2023}
{Zhang} Y., {Adhikari} S., {Costanzi} M., {Frieman} J., {Annis} J., {Chang} C., 2023, The Open Journal of Astrophysics, 6, 46

\bibitem[{{Zu} {et~al}\mbox{.}(2017){Zu}, {Mandelbaum}, {Simet}, {Rozo}, \& {Rykoff}}]{Zu2017}
{Zu} Y., {Mandelbaum} R., {Simet} M., {Rozo} E., {Rykoff} E.~S., 2017, \mnras, 470, 551

\bibitem[{{Z{\"u}rcher} \& {More}(2019)}]{Zurcher2019}
{Z{\"u}rcher} D., {More} S., 2019, \apj, 874, 184

\end{thebibliography}
